\documentclass[sigconf]{acmart}

\AtBeginDocument{%
  }

\usepackage{tikz}
\usepackage{amsmath}
\usepackage{subfig} 

\usepackage{graphicx} 

\usepackage{makecell} 

\usepackage{empheq}
\DeclareMathOperator*{\argmin}{argmin} 
\usepackage{bm} 
\usepackage{url}
\setcopyright{acmlicensed}
\copyrightyear{2024}
\acmYear{2024}
\acmDOI{XXXXXXX.XXXXXXX}

\acmConference[Submitted]{Submitted}{2024}{}
\acmISBN{978-1-4503-XXXX-X/18/06}

\begin{document}


\title{Deep-TEMPEST: Using Deep Learning to Eavesdrop on HDMI from its Unintended Electromagnetic Emanations}

\author{Santiago Fern\'andez}
\email{sfernandez@fing.edu.uy}
\author{Emilio Mart\'inez}
\email{emartinez@fing.edu.uy}
\affiliation{%
  \institution{Facultad de Ingenier\'ia, Universidad de la Rep\'ublica}
  \city{Montevideo}
  \country{Uruguay}
}

\author{Gabriel Varela}
\email{jorge.varela@fing.edu.uy}
\affiliation{%
  \institution{Facultad de Ingenier\'ia, Universidad de la Rep\'ublica}
  \city{Montevideo}
  \country{Uruguay}
}

\author{Pablo Mus\'e}
\email{pmuse@fing.edu.uy}
\author{Federico Larroca}
\email{flarroca@fing.edu.uy}

\affiliation{%
  \institution{Facultad de Ingenier\'ia, Universidad de la Rep\'ublica}
  \city{Montevideo}
  \country{Uruguay}
}


\begin{abstract}
In this work, we address the problem of eavesdropping on digital video displays by analyzing the electromagnetic waves that unintentionally emanate from the cables and connectors, particularly HDMI. This problem is known as TEMPEST. Compared to the analog case (VGA), the digital case is harder due to a 10-bit encoding that results in a much larger bandwidth and non-linear mapping between the observed signal and the pixel's intensity. As a result, eavesdropping systems designed for the analog case obtain unclear and difficult-to-read images when applied to digital video.
The proposed solution is to recast the problem as an inverse problem and train a deep learning module to map the observed electromagnetic signal back to the displayed image. However, this approach still requires a detailed mathematical analysis of the signal, firstly to determine the frequency at which to tune but also to produce training samples without actually needing a real TEMPEST setup. This saves time and avoids the need to obtain these samples, especially if several configurations are being considered. Our focus is on improving the average Character Error Rate in text, and our system improves this rate by over 60 percentage points compared to previous available implementations.
The proposed system is based on widely available Software Defined Radio and is fully open-source, seamlessly integrated into the popular GNU Radio framework. We also share the dataset we generated for training, which comprises both simulated and over 1000 real captures. Finally, we discuss some countermeasures to minimize the potential risk of being eavesdropped by systems designed based on similar principles.

\end{abstract}

\ccsdesc[500]{Security and privacy~Side-channel analysis and countermeasures}
\ccsdesc[500]{Computing methodologies~Neural networks}

\keywords{Software Defined Radio, Side-channel attack, Deep Learning}

\maketitle

\section{Introduction}\label{sec:intro}
TEMPEST is a term used to describe the unintentional emanation of sensitive or confidential information from electrical equipment. 
While it may refer to any kind of emissions, such as acoustic and other types of vibrations~\cite{yu2021keystrokes}, it primarily deals with electromagnetic waves.
In particular, this article focuses on electromagnetic emissions from video displays. The issue of inferring the content displayed on a monitor from the electromagnetic waves emitted by it and its connectors has a long history, dating back to the 1980s with the first public demonstrations by Win van Eck. This problem is sometimes referred to as \emph{Van Eck Phreaking}, but for the remainder of this article, we will use the term TEMPEST~\cite{vaneck1985electromagnetic}.

Van Eck's research was focused on the then-prevalent CRT monitors. However, Markus Kuhn's work in the early 2000s~\cite{kuhn2003compromising} studied modern digital displays, including both the analog interface VGA (Video Graphics Array) and the digital interfaces HDMI (High-Definition Multimedia Interface) or DVI (Digital Visual Interface). Nevertheless, reproducing these studies was challenging due to the need for expensive and specialized hardware, such as a wide-band AM receiver.
This entrance barrier has been significantly reduced in recent years by the development of Software Defined Radio (SDR) ~\cite{wyglinski2016revolutionizing}. SDR employs generic hardware that down-converts the signal to baseband and then provides the sampled signal to the PC, making the hardware more affordable and 
signal processing simpler, since it is performed in software. This advantages resulted in two open-source implementations of TEMPEST (\texttt{TempestSDR} ~\cite{marinov2014remote} and \texttt{gr-tempest}~\cite{larroca2022gr_tempest}) and several empirical studies of the problem, particularly focusing on the HDMI interface~\cite{liu2021screen,lemarchand2020electro,galvis2021denoising,long2024eye,song2015snr,oconnell2019quasi,demeule2020differential,demeule2020quantitative,demeule2020eaves}.  

However, despite all of these efforts ``\textit{this threat still is not well-documented and understood}''~\cite{demeule2020differential}. Our first contribution is precisely to address this issue by providing an analytical expression of the signal's complex samples as received by the SDR when spying on an HDMI display. Virtually all of the above-mentioned studies use an AM demodulation step as part of their processing chain, similar to the first studies by Van Eck with VGA, with the exception of~\cite{demeule2020differential}, which experimentally observed that by using FM demodulation, the attacker may also obtain significant information on the display's content. As we will see, our analytical model explains why both the magnitude and the phase of the complex samples provide information on the eavesdropped image. Furthermore, these expressions are crucial when setting up the eavesdropping system to choose the frequency one should tune to in order to get maximum energy. Instead of tuning the SDR to the frequency that obtains the best Signal-to-Noise Ratio through trial-and-error (as in~\cite{lemarchand2020electro,liu2021screen,long2024eye}), the frequencies to be tested for a particular screen are manageable when based in our analysis.

Equipped with this model, our second contribution is to re-cast the TEMPEST problem as an inverse one. That is, recovering the source image from the baseband complex samples gathered from the SDR. Motivated by the success of deep learning in solving inverse problems in other contexts~\cite{ongie2020inverse}, we propose designing and training a deep convolutional neural network to infer the source image from the baseband complex samples.


To our knowledge, three other works propose deep learning-based algorithms for TEMPEST attacks~\cite{liu2021screen,lemarchand2020electro,galvis2021denoising}. Our work differs significantly, overcoming some limitations of these previous studies. In~\cite{liu2021screen}, the focus is on smartphone displays rather than HDMI or DVI, which emit much lower power signals. They classified almost unintelligible images from \texttt{TempestSDR} into digits, a simpler 10-class classification task. The works in~\cite{lemarchand2020electro} and~\cite{galvis2021denoising} target HDMI but are less applicable to realistic scenarios, processing patches with only a few characters. They both apply a denoiser to the grayscale images produced by \texttt{TempestSDR}. 
Another relevant work is~\cite{long2024eye}, which reconstructs images from electromagnetic emissions of embedded cameras. They used a modified \texttt{TempestSDR} and a GAN-based image translator to restore spied images, offering a potential adaptation to TEMPEST attacks.

\begin{figure}
    \centering   \includegraphics[width=0.45\textwidth]{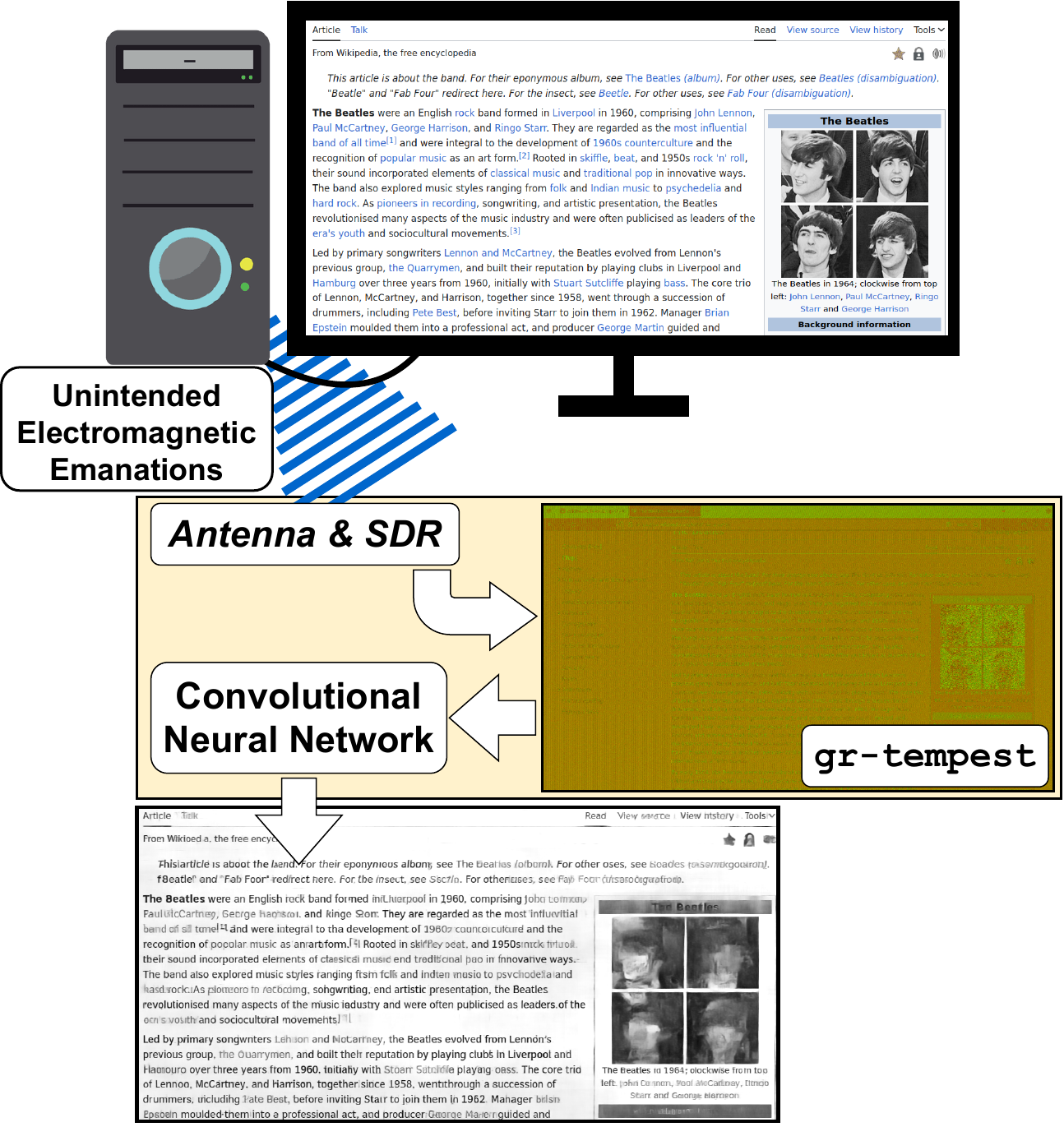}
    \caption{
    Proposed system. The HDMI cable and connectors emit unintended electromagnetic signals, which are captured by the SDR and processed by \texttt{gr-tempest}, obtaining a degraded complex-valued image, which in turn is fed to a convolutional neural network to infer the source image. All three images correspond to actual results.
    }
    \label{fig:sistema_completo}
\end{figure}

More in particular, our contributions in this respect are twofold. Firstly, we have developed and publicly shared an open-source implementation of an end-to-end deep-learning architecture. Figure \ref{fig:sistema_completo} presents an illustrative diagram of the system, including an example of actual results. Our primary focus is on the restoration of text. 
Our architecture surpasses vanilla implementations of either \texttt{TempestSDR} or \texttt{gr-tempest}, producing significantly higher-quality reconstructed images, achieving over 60 percentage points reduction in the average Character Error Rate (CER). Furthermore, and based on the insights provided by our analytical model, we avoid the AM demodulation step all previous works use (as they are based on \texttt{TempestSDR}), which further distorts the signal, and instead learn to map directly from the complex samples to the original image; i.e.\ solve the inverse problem. As we report in Sec.\ \ref{sec:results}, using the complex samples and avoiding the information loss incurred in demodulation results in a significant gain in performance. 


Secondly, we have made this article's complete dataset publicly available. It includes two sources of data: several real-life signals and a GNU Radio-based simulator, which we developed and are sharing, that, given an image, produces the spied signal. This simulator is based on the analytical expressions derived in this work. Furthermore, we discuss how to train the learning module (partially) based on these simulations, significantly reducing the time-consuming stage of acquiring real-life signals without negatively impacting the quality of the recovered images. The full dataset comprises around 3500 samples, out of which approximately 1300 are real captures. Our aim is to make this openness useful in further advancing research in this area. Please visit \url{https://github.com/emidan19/deep-tempest} for the complete dataset and code.

The rest of the article is structured as follows. The next section discusses the threat model, whereas Sec.\ \ref{sec:hdmi} provides a detailed overview of the HDMI signal. In Sec.~\ref{sec:sdr}, we summarize the working principle of SDR and characterize the forward operator by giving a mathematical expression of the samples produced by the hardware given an input image. How to recover the image from these samples by means of deep learning is discussed in Sec.~\ref{sec:dl}. The obtained results and countermeasures are presented in Secs.~\ref{sec:results} and~\ref{sec:countermeasures_robustness}. Closing remarks and future work are discussed in Sec.~\ref{sec:conclusions}.


\section{Threat Model}\label{sec:threat}

This section presents the threat model we consider in this work. The attacker's objective is to recover the image displayed on a monitor that contains sensitive or confidential information. This monitor is connected through a standard digital display interface, which may be either HDMI or DVI. To achieve their objective, the attacker will resort to the electromagnetic energy emanating from the connectors and cables of the digital display, from which they will infer the monitor's content. 

We assume that the attacker is equipped with off-the-shelf hardware to capture and process these emanations. The necessary equipment includes a laptop with a GPU (although a CPU-only laptop is a viable, albeit slower, alternative), an SDR hardware (see Sec.\ \ref{sec:sdr} for a discussion), an antenna, and a Low Noise Amplifier (LNA). 

We foresee two separate operational scenarios. Firstly, one where the attacker remains unnoticed, e.g., if the spied system is close to a wall and the attacker operates from the other side. In this case, the setup may include somewhat large directive antennas, and an online operation is viable where, for instance, the attacker adjusts the antenna's direction until a proper image is obtained and only saves the images that they are interested in.

A second scenario is one where only the attacker's hardware goes unnoticed. For instance, a small omnidirectional antenna is left near the HDMI cable and connectors of the spied system, and the spying PC is not visible or does not draw attention. In this case, which requires physical proximity to the spied system, the attacker's PC may periodically (e.g., \ every second) record a signal, process it to obtain an image, and save it for offline visualization. If hard drive space is not an issue, the attacker may even record the raw samples of the SDR periodically and apply our method to these recordings. 

\section{Unintended Electromagnetic Emanations of HDMI}\label{sec:hdmi}

\subsection{Digital signal}

Although there are seven different versions of HDMI (ranging from 1.0 up to 2.1) and five types of connectors (A to E), video is encoded the same way for all of them except for version 2.1. This last version, released in 2017, is typically used only in high-end TVs with 4k or 8k video, and we will not consider it in this work. In any case, HDMI is backward compatible with single-link DVI, so our results are also valid for DVI-D or DVI-I. 

\begin{figure}
    \centering
    \includegraphics[width=0.45\textwidth]{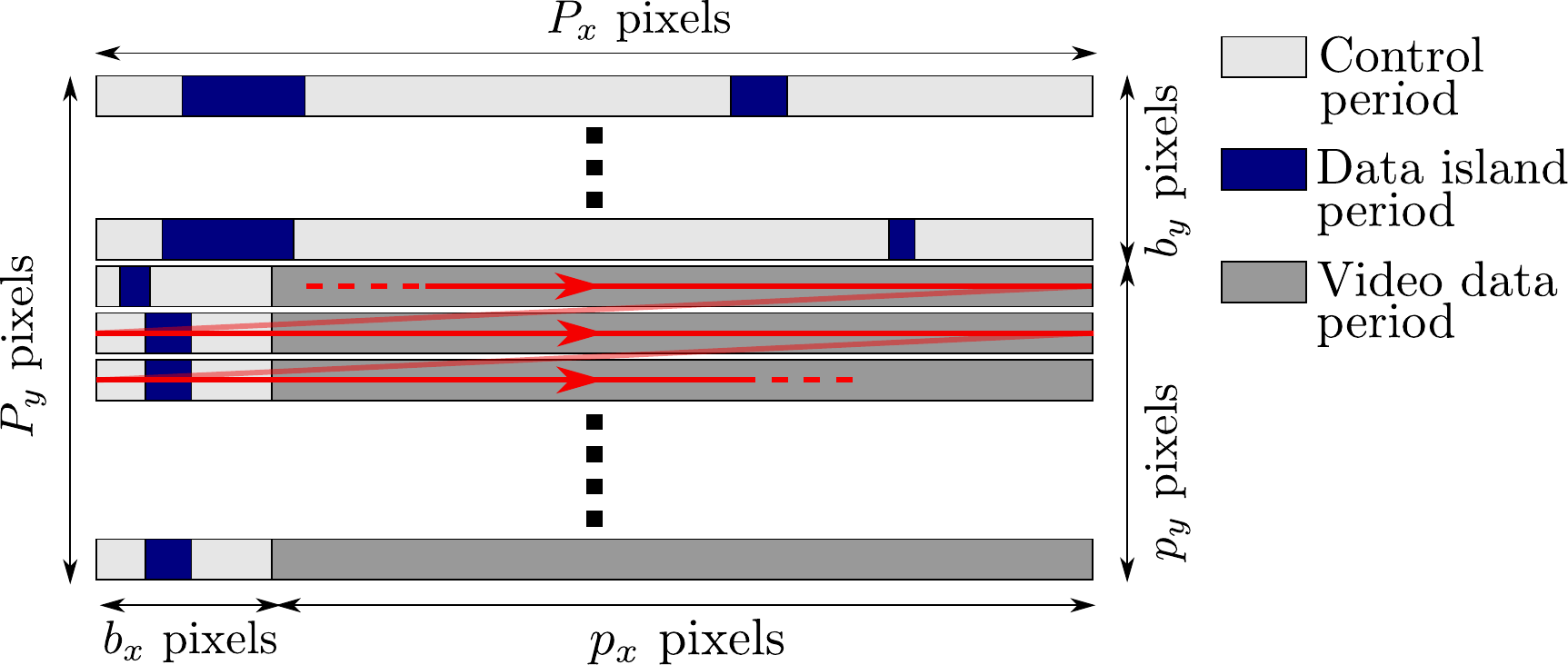}
    \caption{An illustration of the transmission of a frame on a single TMDS channel. The red arrow indicates the order in which the signal is transmitted. Video is actually sent only during the video data periods. }
    \label{fig:frame}
\end{figure}

To transmit audio and video, HDMI uses three separate TMDS channels, each corresponding to the red, blue, and green components regarding video, where each channel is sent serially over three separate pins (positive, negative, and ground; further details regarding the electrical signal are presented in the next subsection). While $YC_bC_r$ pixel encoding and other color depths are possible, the default configuration is $RGB$ encoding with 24 bits. We will thus only consider this configuration for brevity, although extensions to these scenarios are straightforward. As illustrated in Fig.\ \ref{fig:frame}, and just as in VGA, each video frame includes a horizontal and vertical blanking, where no video is transmitted. During these periods, audio or control packets are transmitted instead (the so-called control and data island periods). 

This means that the pixel rate is actually higher than what is being displayed. For instance, for a resolution of $1920\times 1080$ with progressive scan, there are actually $2200\times 1125$ pixels per frame (including blanking). In terms of the notation of Fig.~\ref{fig:frame}, this means that $p_x=1920$, $p_y=1080$, $P_x=2200$ and $P_y=1125$, which at a frame rate of 60~Hz represents a pixel rate of $1/T_p=148.5$~MHz. Supported resolutions and the corresponding timings may be consulted at the EIA/CEA-861 standard, but it is important to note that the possibilities are limited (e.g.~197 possible timings and resolutions in HDMI 2.0, and only 64 for HDMI 1.4). 


Different from VGA, the intensity of each color (from 256 possible values) is encoded into 10 bits before transmission. 
The 8-bit input word is first differentially XORed or XNORed using the first bit as the reference. The encoder uses the operation that results in fewer bit transitions given the input word, and the choice is indicated in the ninth bit. The second stage negates or not the first 8 bits (flagged by the tenth bit) to even out 1s and 0s in the encoded stream. Note that each video data period is encoded independently, meaning that the process is restarted for each line.

\subsection{Electrical and electromagnetic signal}

After analyzing the digital signal generated by the video, we can now examine the resulting electromagnetic signal surrounding the cable. Our main interest is to determine where the largest portion of its power lies in the spectrum so we can tune our system to that frequency. Additionally, we want to obtain an approximate expression of this electromagnetic signal, which will help us simulate it. This will enable us to produce samples that we can use to train and evaluate our learning system without necessarily using an actual TEMPEST setup. We will defer this last problem to the next section since it also includes the effects of the SDR hardware.

HDMI uses differential signaling, basically meaning that every channel is composed of two cables, where the bit value is estimated from the difference in voltage between the two. That is to say, for any of the three TMDS channels, the voltage signal $x^+(t)$ and $x^-(t)$ in both cables would be:
\begin{gather}
    x^+(t) = V_{cc} + \sum_k x_b[k]p(t-kT_b),\\
    x^-(t) =V_{cc}- \sum_k x_b[k]p(t-kT_b),
\end{gather}
where $V_{cc}$ is a constant, $x_b[k]$ corresponds to the mapping of $k$-th bit (e.g.\ a negative voltage for 0 and a positive one for 1), $T_b$ is the bit duration, and $p(t)$ is the shaping pulse (typically a rectangular pulse of duration $T_b$). 

The immediate consequence is that under an ideal system and observing both cables together as in our case, we would measure  $x(t)=x^+(t) + x^-(t) = 2V_{cc}, $ which is independent of the information-carrying sequence $x_b[k]$. However, as observed in previous works~\cite{song2015snr}, the pulses in $x^+(t)$ and $x^-(t)$ are not perfectly aligned nor exactly the same. For instance, assuming that $x^-(t)$ is delayed a time $\epsilon T_b$ with respect to $x^+(t)$, we would obtain
\begin{align}\label{eq:pcm_posta}
    x(t) =  x^+(t) + x^-(t) = & 2V_{cc} + \sum_k x_b[k]q(t-kT_b),\\
    \text{where }q(t) = & p(t)-p(t-\epsilon T_b).\label{eq:ejemplo_q}
\end{align}

That is to say, ignoring the constant $2V_{cc}$, a classic PCM (Pulse-Code Modulation) signal with conforming pulse $q(t)$. By adding a random delay to $x(t)$, we can study it as a Wide-Sense Stationary signal whose Power Spectral Density (i.e.\ the expected power per Hertz) has the following well-known expression:
\begin{gather}\label{eq:psd_posta}
    S_X(f) = \frac{|Q(f)|^2}{T_b}S_{X_b}(f) = \frac{4\sin^2{(\pi f \epsilon T_b)}}{T_b}\text{sinc}^2(fT_b)S_{X_b}(f),
\end{gather}
where $S_{X_b}(f) = \sum_l R_{X_b}[l] e^{-j2\pi fl T_b}$ and $R_{X_b}[l] = \mathbb{E}\{x_b[k]x_b[k+l]\}$. That is to say, the Discrete-Time Fourier Transform $S_{X_b}(\omega)$ of the auto-correlation of the sequence $x_b[k]$ evaluated at $\omega=2\pi fT_b$. Note that $S_{X_b}(f)$ is a periodic function of period $1/T_b$ (the bit rate).

It is typically the case that consecutive frames in the spied monitor are very similar (if not identical). This is also true for contiguous lines. Denoting as $T_p$ the pixel time (i.e.\ $T_p=10T_b$), and recalling that each line is encoded independently, the previous two observations mean that high values of $S_{X_b}(f)$ should be expected at multiples of $f=1/(P_xP_yT_p)$ (the frame rate) as well as $f=1/(P_xT_p)$ (the horizontal lines rate). Furthermore, given that TMDS encoding enforces no DC component, $S_{X_b}(0)\approx 0$. 

The other relevant time scale is precisely $T_p$ since consecutive pixels are similar. Note that the analysis in this case is complicated by the non-linear encoding we discussed before. As a first step, let us consider a constant image, which produces at most two different encoded words (the differentially encoded word or its negation), which are sent alternately, the least significant bit first. This process will produce a $S_{X_b}(f)$ with large spikes at every multiple of $1/T_p$ since under a constant image, bits 10-bits apart are typically the opposite (i.e.\ typically $x_b[k]=-x_b[k+10]$). Another significant spike should be present at $1/(2T_p)$, too, since bits 20-bits apart are typically the same. 

\begin{figure}
    \centering
    \includegraphics[width=0.45\textwidth]{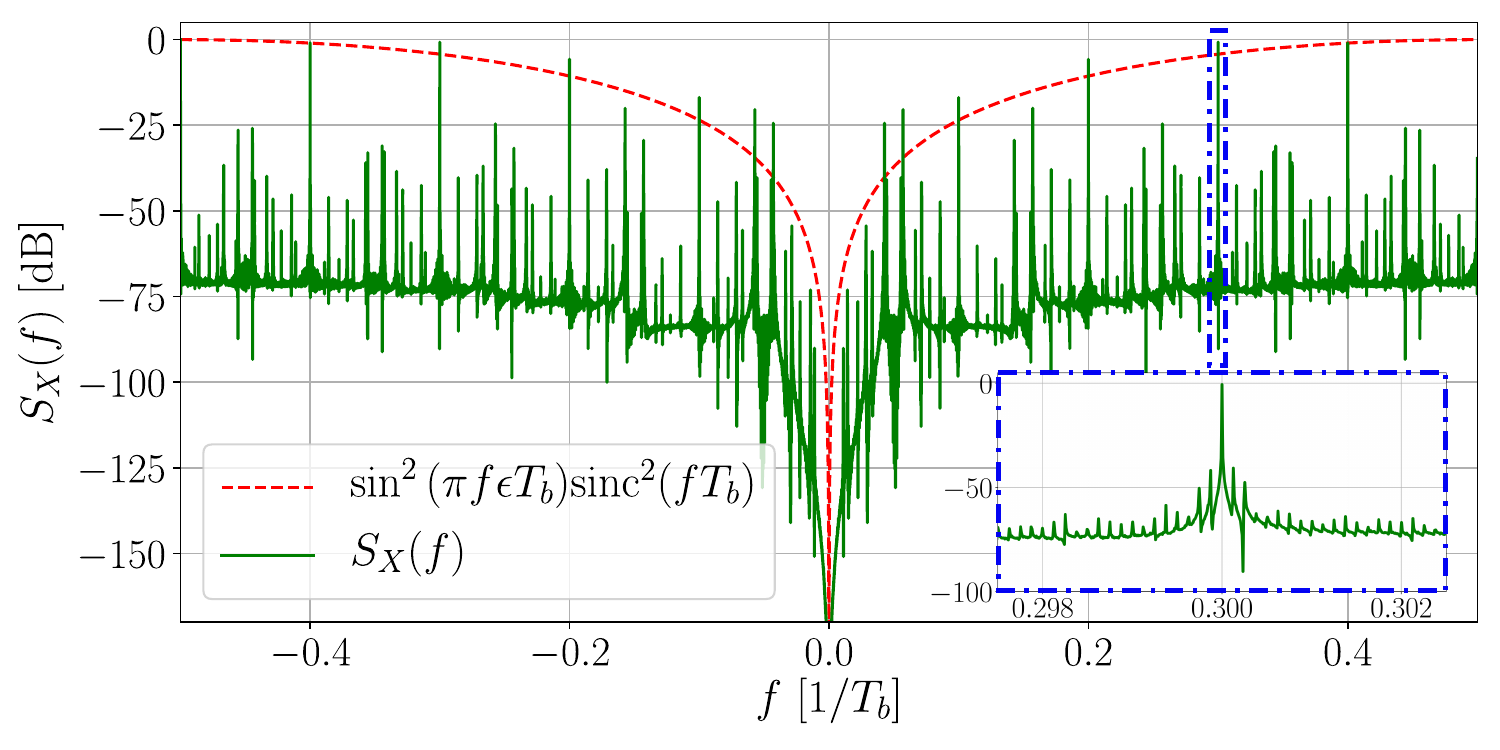}
    \caption{
The power spectral density of a TMDS encoded signal computed by multiplying an estimation of $S_{X_b}(f)$ and $|Q(f)|^2/T_b$ (the dashed red curve, shown for reference); cf.\ Eq.\ \eqref{eq:psd_posta}. Both curves are normalized to its maximum value for clarity. Significant spikes every multiple of $0.1/T_b$ are clearly visible. In the zoom-in around $f=0.3/T_b$ shown below, smaller but nevertheless important spikes every multiple of $1/(P_xT_p)$ (the inverse of the duration of each horizontal line) are also clearly visible.
    }
    \label{fig:psd_final}
\end{figure}

This intuition is verified for more complex encoded images, as shown in Fig.~\ref{fig:psd_final}, which displays an estimation of $S_{X_b}(f)$ for a TMDS signal corresponding to eight frames of a user typing in a word processor, multiplied by $|Q(f)|^2/T_b$ (cf.\ Eq.\ \eqref{eq:psd_posta}) along with $|Q(f)|^2$ for reference  (using $\epsilon=0.002$). Note that the significant increase in $S_{X_b}(f)$ at $f\approx 0.05/T_b=1/(2T_p)$ is attenuated by $|Q(f)|^2$, whereas the peaks every multiple of $0.1/T_b=1/T_p$ are not. The lower graph in the figure displays a zoom-in to the third-pixel harmonic (marked with a blue slashed rectangle), where the peaks corresponding to multiples of $1/(P_xT_p)$ are clearly visible.

The conclusion of this section is that most of the power of the emanations from an HDMI signal is located at the first few multiples of the pixel rate. Naturally, the precise expression of $q(t)$ in \eqref{eq:pcm_posta} is not known a priori. In \eqref{eq:psd_posta}, we have only assumed unaligned pulses (with an unknown $\epsilon$), but other differences may also exist. Regarding where most of the leaked power exists, a first approximation, like the one we presented, is enough. Furthermore and quite interestingly, as discussed in the following two sections, this expression will also be enough to produce simulations that may be used to train a learning system that maps samples of the emitted signal to the source image that produced them. 




\section{Software Defined Radio}\label{sec:sdr}

Having characterized our signal of interest $x(t)$ in \eqref{eq:pcm_posta}, let us now discuss how to intercept it and, furthermore, provide an analytic expression to the signal captured by the SDR and thus the one we may consider to perform the eavesdropping.  

\subsection{Hardware}

\begin{figure}
    \centering
    \includegraphics[width=\linewidth]{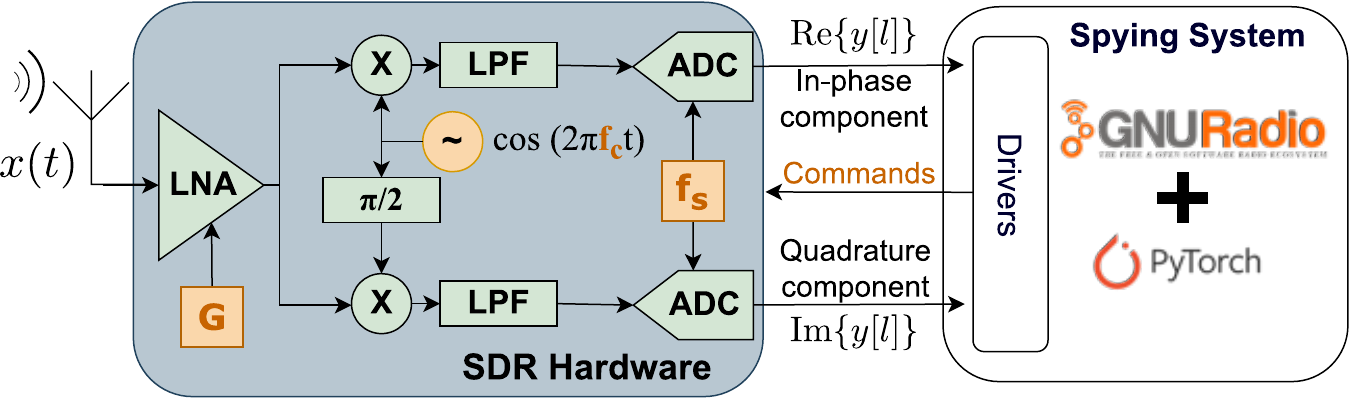}
    \caption{Diagram of an SDR. 
    The drivers provide complex samples $y[l]$ whose real and imaginary parts correspond to the in-phase and quadrature components. }
    \label{fig:SDR_and_tempest}
\end{figure}

\begin{figure}
    \centering
    \includegraphics[width=0.35\textwidth]{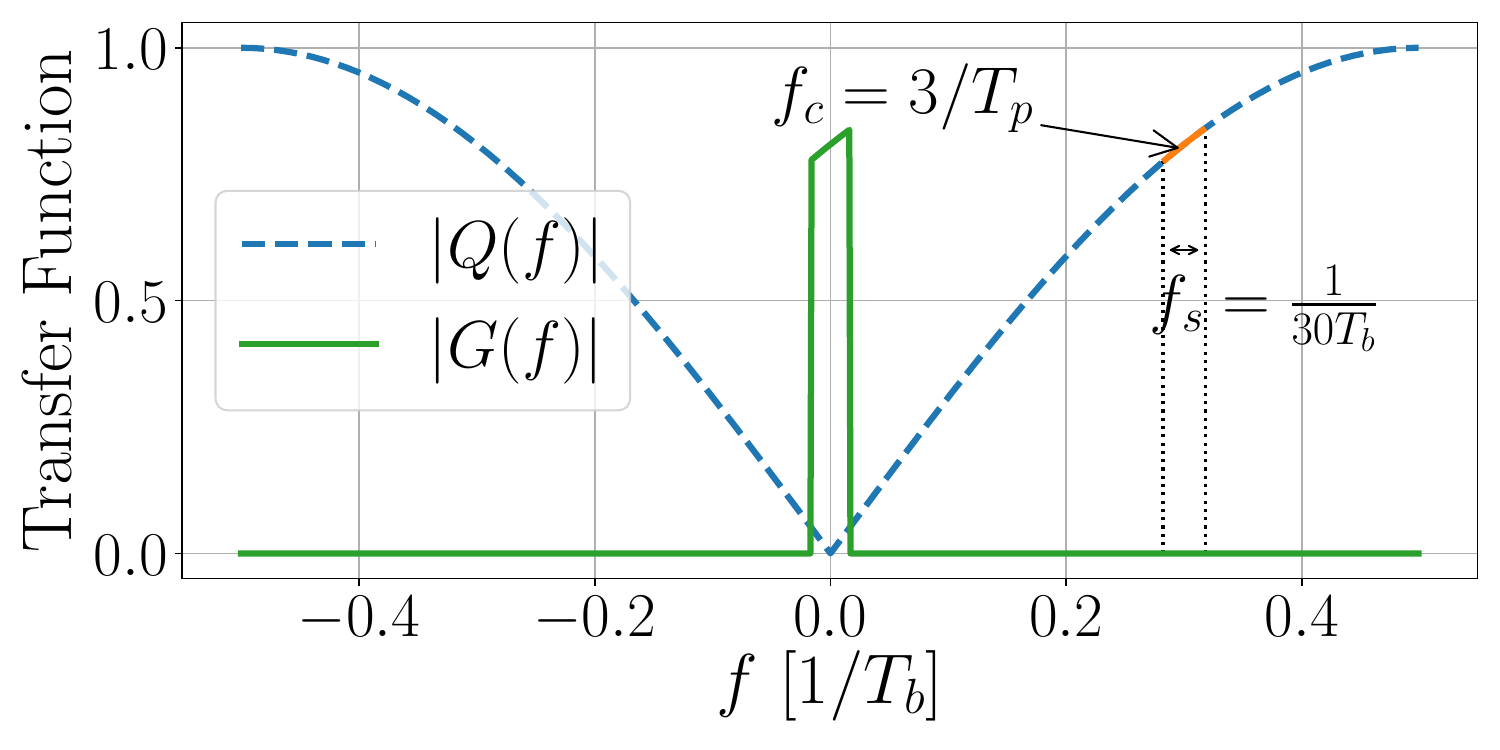}
        \caption{Normalized Fourier Transform of $q(t)$ (i.e.\ Eq.\ \ref{eq:ejemplo_q} with $\epsilon=0.002$) and $g(t)$, the complex baseband representation of the channel as seen by the SDR. 
        }
    \label{fig:ejemplo_G}
\end{figure}

As illustrated in Fig.~\ref{fig:SDR_and_tempest}, an SDR hardware moves the signal to baseband and provides its filtered samples. These samples will be processed using software to produce the eavesdropped image. 
Starting from~\eqref{eq:pcm_posta}, and ignoring the constant term, we may interpret $x(t)$ as a train of Dirac deltas that goes through a filter with impulse response $q(t)$. However, since we are down-converting this signal to baseband, the complex baseband representation of this channel is actually a filter with impulse response $g(t)=\mathcal{F}^{-1}\{Q(f+f_c)H_{LFP}(f)\}$ (see for example~\cite{gallager2008principles}). That is to say, the inverse Fourier transform of the product between the Fourier transform of $q(t)$ moved to zero from the tuning frequency $f_c$ (which, as we discussed before, will be equal to a harmonic of $1/T_p$) times the transfer function of the SDR's low-pass filter. If a sampling rate $f_s$ is used, then $H_{LPF}(f)$ is ideally zero for $|f|>f_s/2$ and a constant otherwise. In other words, instead of filtering the train of Dirac deltas with $q(t)$, we use $g(t)$, whose Fourier transform $G(f)$ is $Q(f)$ evaluated around $f_c$ and zeroed for $|f|>f_s/2$. This process is illustrated in Fig.~\ref{fig:ejemplo_G} using $q(t)$ as defined in~\eqref{eq:ejemplo_q}, $f_c=3/T_p$ and $f_s=1/(30T_b)$.

All in all, after sampling, the following sequence is obtained:
\begin{gather}\label{eq:senial_capturada}
    y[l] = \sum_k x_b[k]g(l/f_s - kT_b).
\end{gather}
We may further enrich the model by adding noise, small errors to $f_c$ (instead of precisely a multiple of the pixel rate), and offsets in both time and phase (uniform between zero and $1/f_s$ or $2\pi$, respectively). These impairments are included in our simulations to make the learning system more robust to these non-idealities. Note, however, that we are ignoring the antenna's bandwidth and possible non-linearities.

Regarding the sampling rate, mid-level SDRs allow for, at most, some tens of MHz. For example, the USRP 200-mini~\cite{USRPmini} we used in our experiments has a maximum sampling rate of $f_s=50$~MHz. Just as in the example in Fig.~\ref{fig:ejemplo_G}, this is only a third of the pixel rate at a resolution of $1920\times 1080$@60Hz (resulting in $1/T_p= 148$~MHz), meaning that each sample $y[l]$ will actually be a linear combination of several tens of encoded bits, further complicating the image reconstruction. 

In fact, since the anti-aliasing filter of the SDR produces a $G(f)$ that is zero for $|f|>f_s/2$, and if $f_s\ll 1/T_b$ as we just discussed, the resulting loss of information means that the attacker cannot recover the sequence of bits $x_b[k]$ by observing the samples $y[l]$.
It may appear that a viable alternative is to increase the sampling rate $f_s$ up to $1/T_b$, and after equalization, sample each bit separately and decode the image. There are three important drawbacks to this approach. Firstly, it would require an SDR that operates with a sampling rate and a corresponding instantaneous bandwidth of at least some GHz, which even high-end and extremely expensive solutions struggle to provide (e.g.\ the USRP X440 by Ettus Research provides up to 3200 MHz of bandwidth at the cost of over 25,000 dollars~\cite{x440}). Secondly, it is unclear if the interference from other sources (received due to the increased receiver's bandwidth) will not prove detrimental in recovering the image. Last but not least, there is the problem of processing such an enormous amount of samples, which would further impact the resulting cost of the spying setup, this time in terms of the required PC. 

For the above reasons, we will consider a sampling rate value $f_s$ as those obtained from less expensive (and also less conspicuous) hardware, which will thus unavoidably result in an unrecoverable bit sequence $x_b[k]$. However, recall that the attacker's actual objective, as in any communications problem, is to estimate the most plausible image that generated the observed complex sequence $y[l]$. 
We propose a data-driven approach to this problem that leverages the \emph{a priori} information regarding what kind of images are typically displayed in a monitor (i.e., \ the original images used in the training set should be representative of desktop content). This is accomplished through a deep-learning module, which we present in detail in the next section. 
Before that, the following subsection discusses how, for the sake of simplicity, this estimation is simply computed as $|y[l]|$ in \texttt{TempestSDR}. 
\subsection{Software}
\label{subsec:software}

Regarding software, samples are provided by the driver and then processed arbitrarily by the spying PC. Both \texttt{TempestSDR} and \texttt{gr-tempest} adapt the sampling rate $f_s$ to produce an integer number of samples for every $P_x$ pixels, i.e.,\ $P_xT_p=m/f_s$ for some integer $m$. When the sampling rate is successfully synchronized this way, these $m$ samples correspond to a line, and thus, displaying $P_y$ of these lines produces a non-skewed and static image. Correlations as the one we discussed before are searched for in the signal and used in a PLL-like system to estimate the precise value of $f_s$ (see \cite{larroca2022gr_tempest} and \cite{marinov2014remote} for details). 

Given that \eqref{eq:senial_capturada} is a complex signal (as seen in Fig.~\ref{fig:ejemplo_G}, since $|G(f)|$ is not symmetric around zero), \texttt{TempestSDR} actually takes the magnitude of the samples (i.e.\ an envelope detector, termed AM demodulator in some contexts, e.g.~\cite{long2024eye}), which further distorts the signal. To avoid this unnecessary degradation, for the case of VGA \texttt{gr-tempest} instead applies an equalization filter to the complex signal to produce much better results. We will also consider the complex signal so as to provide the learning system with the most information available. As we will see, this choice will have a non-negligible impact on the performance of the model.

The other significant difference between \texttt{TempestSDR} and \texttt{gr-tempest} is that the former was coded from scratch, whereas the latter uses GNU Radio~\cite{gnuradio}. This is a framework that represents a processing chain as a series of interconnected blocks (a so-called \emph{flowgraph}), each executing a well-defined operation on the signal (e.g.\ filtering or resampling). New blocks can be easily created and added to the already vast list of available ones. These new blocks can be programmed either in C++ or Python. In the latter case, Numpy is used to represent data, which further simplifies the integration of deep learning frameworks such as PyTorch, as in our case. All of these features have been the main motivation behind our choice of \texttt{gr-tempest} as the starting point of our system. 



\section{Eavesdropping Images from \texttt{gr-tempest} Complex Sequences}\label{sec:dl}

\subsection{Deep Learning to Solve the Inverse Problem}



In this section, we consider the inverse problem of recovering a clean or source image $\bm{X} \in \mathbb{R}^{p_y \times p_x}$ from a degraded observation $\bm{Y} \in \mathbb{C}^{p_y \times p_x}$, which is an array of complex numbers with equal size of the source image. This observation is modeled as: 
\begin{equation}
    \bm{Y} = \mathcal{T}(\bm{X}) + \bm{N},  
\end{equation}
where $\mathcal{T}\colon \mathbb{R}^{p_y \times p_x} \to \mathbb{C}^{p_y \times p_x}$ is a non-linear degradation operator, and $\bm{N} \in \mathbb{C}^{p_y \times p_x}$ is an additive complex noise, for which real and imaginary parts are assumed to be mutually independent, each of them being a white Gaussian noise image of variance $\sigma^2$. Recall that in our case, $\bm{X}$ refers to a monitor image to be spied on (and thus of shape $p_y\times p_x$), while $\bm{Y}$ corresponds to an array of complex samples defined by~\eqref{eq:senial_capturada} and synchronized by \texttt{gr-tempest}. More details on how we construct $\bm{X}$ and $\bm{Y}$ are discussed in the following subsection.

Due to the aforementioned inter-symbol interference, the degradation operator $\mathcal{T}$ is severely ill-posed, so achieving perfect restoration of $\bm{X}$ is impossible. Therefore, we must settle for obtaining an estimation $\hat{\bm{X}}$ by introducing regularization and hope to get as close as possible to the original image. This corresponds to performing Bayesian estimation to solve a Maximum A Posteriori problem, which can be formulated as follows:
\begin{equation}
    \hat{\bm{X}} = \argmin_{\bm{X}} \frac{1}{2\sigma^2}\|\bm{Y} - \mathcal{T}(\bm{X})\|^2 + \lambda \mathcal{R}(\bm{X}),
    \label{map_equation}
\end{equation}
where the solution minimizes a data term $\frac{1}{2\sigma^2}\|\bm{Y} - \mathcal{T}(\bm{X})\|^2$ and a regularization term $\lambda \mathcal{R}(\bm{X})$ with regularization parameter $\lambda$. Specifically, the data term is responsible for demanding similarity with the degradation process, while the regularization term is composed of a function $\mathcal{R}\colon \mathbb{R}^{p_y \times p_x} \to \mathbb{R}_+$ that holds responsibility for delivering a stable solution. The proper choice of a regularizer is not a trivial task as it involves considering prior knowledge of the kind of images to be recovered. 
However, traditional hand-crafted priors (e.g.\ Tikhonov regularization) are usually too over-simplistic and do not capture the complexity of real images. This is why recent methods follow learning-based approaches that, using large datasets of pairs of source/degraded image samples, directly learn the mapping from the degraded observations to the source images~\cite{zhang2018ffdnet}
or learn decoupled priors combined with the MAP formulation~\cite{zhang2017learning}.

In this work, we propose to train an end-to-end deep convolutional neural network (CNN) as a regressor $\hat{\bm{X}} = f(\bm{Y},\Theta)$ to learn to map the degraded complex signals, spied, into the clean source images. This training is performed by minimizing a certain loss function $\mathcal{L}$ on a training set containing $N$ clean-degraded image pairs $\left\{(\bm{X}_i, \bm{Y}_i) \right\}_{i=1}^{N}$, i.e. 
\begin{equation}
    \min_{\Theta} \sum_{i=1}^{N} \mathcal{L} \bigl( f({\bm{Y}_i}, \Theta), \bm{X}_i \bigl). 
        \label{arg_min_cost_problem}
\end{equation}
Note that the CNN regressor $f({\bm{Y}_i}, \Theta)$ does not depend on the degradation operator $\mathcal{T}$ explicitly, but it does so in an implicit way since the clean-degraded image pairs that are used to compute its weights in~\eqref{arg_min_cost_problem} may be synthetically generated \emph{via} $\bm{Y}_i =\mathcal{T}(\bm{X}_i)$. 

\begin{figure}
    \centering
    \includegraphics[width=\linewidth]{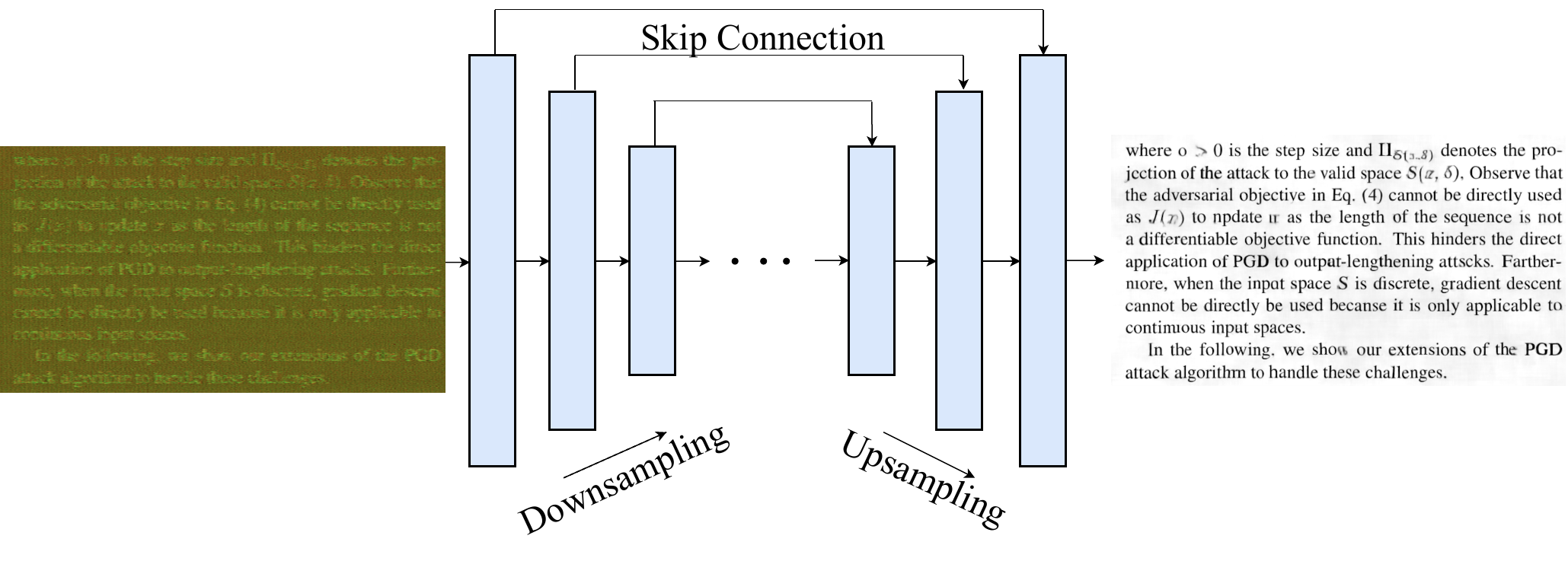}
    \caption{DRUNet architecture takes as input the in-phase and quadrature components (red and green channels, respectively) of the eavesdropped image and outputs a grayscale image.
    }
    \label{fig:drunet}
\end{figure}

For the network $f({\bm{Y}_i}, \Theta)$ we use DRUNet (\textit{Deep Residual UNet})~\cite{PlugAndPlayDRUnet}, a popular CNN with high expressive power. Its architecture, depicted in Fig.~\ref{fig:drunet}, is composed of a succession of interconnected convolutional layers, activation functions, and pooling or subsampling layers. Inspired by UNet~\cite{UNet}, DRUNet uses an encoder-decoder structure: in the first series of convolutional layers, the image is down-sampled to a lower-dimensional space, and then, throughout the second series of convolutional layers, the image is up-sampled to its original size. 
Furthermore, as in other architectures like ResNet \cite{ResNet}, it is possible to interconnect non-adjacent convolutional layers using residual blocks and \textit{skip connections}. This strategy has been shown to enhance the model capacity.

\subsection{Generating the training set}

\begin{figure}
    \centering
    \includegraphics[width=\linewidth]{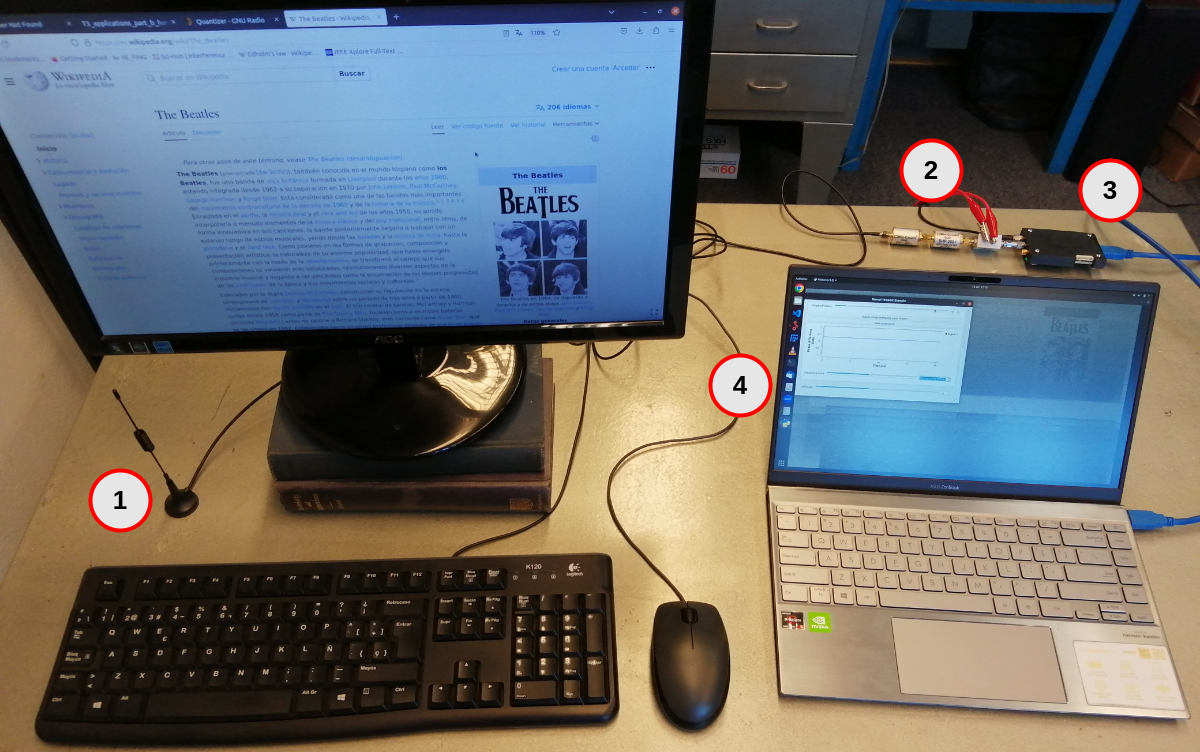}
    \caption{Experimental setup. The enumeration corresponds to 1) antenna, 2) RF filters and amplifier, 3) SDR, and 4) the spying computer running a GNU Radio flowgraph. 
    }
    \label{fig:exp_setup}
\end{figure}

Let us now discuss how we constructed the training set. Each pair $(\bm{X}_i,\bm{Y}_i)$ stems from two possible sources: actual spied signals or simulations. The former was obtained using the experimental setup shown in Fig.~\ref{fig:exp_setup}. The antenna was placed somewhat close to the cable and complemented with a Mini-Circuits ZJL-6G+ amplifier and a band-pass filter composed of an SLP-450+ low-pass filter and an SHP-250+ high-pass filter, both from Mini-Circuits. This would correspond to the second scenario we discussed in Sec.\ \ref{sec:threat}. It is worth mentioning anyhow that we are not interested in proving the feasibility of TEMPEST, which has already been demonstrated~\cite{marinov2014remote,kuhn2003compromising,song2015snr,oconnell2019quasi}, but in improving the results obtained by the state of the art (i.e.\ when \texttt{gr-tempest} or \texttt{TempestSDR} obtain reasonable results, our system should further improve them). Our simple setup is sufficient to this end. 

It is important to emphasize that obtaining real captures is not a simple task. 
We used a monitor with a resolution of $1600\times900$ @ 60 fps, tuning the SDR to the third harmonic of the pixel frequency (324~MHz) using a modified version of the flowgraph of \texttt{gr-tempest}. Modifications include minor improvements to the tuning frequency correction algorithm, and naturally output the complex samples instead of their magnitudes. Furthermore, as we mentioned before, \texttt{gr-tempest} automatically adapts the sampling rate $f_s$ to produce an integer number $m$ of samples per image row. We have further interpolated this signal to produce $P_x$ complex samples per line (i.e.\ using an interpolator with ratio $P_x/m$).

The most challenging aspect was tagging which sample corresponds to the first pixel of the image. 
%
This is a key step in performing a pixel-by-pixel matching of the captured image with its respective original version, which is necessary for the model's supervised training. 
Although \texttt{gr-tempest}, in addition to adapting the sampling rate $f_s$, also provides an automatic algorithm that re-centers the image, in our experience, the results of the latter were not sufficient for our purposes. 

To address this limitation, we first detect the blanking periods using the Hough Line Transform~\cite{hough_transform_survey}. We then both remove them entirely and shift the image, leaving the capture adjusted to the original version. Detection is achieved by keeping only those lines whose distance between each other corresponds to the blanking size (for both horizontal and vertical). 
A grayscale conversion of the original image constitutes $\bm{X}_i$ (more in particular, the average of the three \textit{RGB} image color channels), whereas the re-centered complex array of the samples 
constitutes the corresponding $\bm{Y}_i$.  


The rest of the degraded images were simulated under the same conditions as the SDR (sampling rate and tuning frequency) and the system being eavesdropped (resolution). The synthetic dataset was generated with a Python script, also available at the project's repository, that simulates the pipeline composed of the HDMI transmission protocol, the SDR baseband down-conversion, and low-pass filtering and sampling (i.e.\ Eq.~\eqref{eq:senial_capturada}). Gaussian noise, small frequency errors, and a random delay were also added. To explore the effects of using a precise expression of the pulse $q(t)$, we have tested two different possibilities: the difference between two delayed rectangular pulses (as in~\eqref{eq:ejemplo_q} with $\epsilon=0.1$), or simply a rectangular pulse. As we will see, quite interestingly, the trained system is robust to this choice.

\section{Experiments and Results}\label{sec:results}



We gathered a set of $3491$ clean-degraded image pairs following the procedure presented in the previous section. 
The dataset includes 2189 simulated samples for each pulse (1738 used for training, 148 for validation, and 303 for test) as well as 1302 real-life samples (882 for training, 120 for validation, and 300 for test).
The dataset was carefully constructed to represent the content of an actual screen image, ranging from online sales pages~\cite{kumar-etal-2022-cova} to conference articles~\cite{CVPR2019} and manual screenshots on a variety of web pages.



To evaluate the performance of the trained models, we first need to define a representative restoration metric. Typical image restoration metrics are the Peak Signal-to-Noise Ratio (PSNR) or the Structural Similarity Index Measure (SSIM)~\cite{pedersen2012full}. However, it is reasonable to assume that the eavesdropper is mostly interested in the text being displayed on the monitor. In this case, neither of them are suitable indicators as they are sensitive to changes in the images' contrast and are thus not indicative of the legibility of the recovered text. 
For this reason, we chose to also report the Character Error Rate (CER), which was computed using the Tesseract optical character recognition software~\cite{TessOverview}. We remark that the OCR system was only used to evaluate performance, not for model training. In particular, we compare the text produced by Tesseract on the original image and on the recovered one. The percentage of different characters between both outputs is the CER, and we report the average over all images in the test set. 

The hardware used for training and evaluation tasks consists of an Intel Core i7-10700F CPU with 64GB of RAM and an NVIDIA GeForce RTX 3090 GPU with 24GB of VRAM. Inference on $1600\times900$ sized images takes approximately $0.5$s with GPU and $15$s on CPU.
The model parameters were optimized by minimizing the $L2$ norm between the recovered image and its ground truth. We used the Adam optimizer~\cite{kingma2017adam} to train on image patches of $256 \times 256$ pixels (\textit{patch size}) and batches of 48 patches (\textit{batch size}). A Total Variation regularizer~\cite{chan2006total} was also added to reduce noise while preserving the edges. The values of the learning rate ($lr=1.56\times10^{-5}$) and the regularization weight ($\lambda_{TV}=2.2\times10^{-13}$) were found through a hyper-parameter search using the Optuna framework~\cite{akiba2019optuna}. Weights of the DRUNet architecture were initialized with He's  Normal weights~\cite{he2015delving}, except for certain cases we discuss below.


\noindent \textbf{Synthetic data only.} Let us first consider an ideal case where we perfectly know the electromagnetic signal's behavior, i.e.\ a model trained and evaluated only on the synthetic data. We shall denote it as \emph{Base Model}, and it will be useful both to assess the impact of the approximations we performed when deriving \eqref{eq:senial_capturada}, but also to evaluate what performance we may expect (at best) when using real-life signals. 
As we mentioned in the previous section, we have trained and evaluated our system using two different pulses: a rectangular pulse, or a difference of two rectangular pulses as in~\eqref{eq:ejemplo_q}, with $\epsilon=0.1$. We trained both models 180 epochs, resulting in a CER of around 30\% when tested over their respective synthetic samples. The complete set of results is summarized in Table \ref{table:evaluacion_test_metodos}. 

\noindent \textbf{Evaluation in real-life data.} Next, we consider real-life signals acquired with the setup displayed in Fig.~\ref{fig:exp_setup}. If we evaluate both Base Models on this data, their performance drops significantly to a CER of about 50\%, still much better than those of the grayscale images produced by both \texttt{TempestSDR} or vanilla \texttt{gr-tempest}, which obtain a CER of over 90\%. Furthermore, the fact that both Base Models obtain similar results indicates that a precise expression for the conforming pulse $q(t)$ is unnecessary, which we will further explore in the next section.  However, synthetic data will prove significantly useful when combined with real-life signals, dramatically decreasing the number of samples required in training, a discussion we defer to the end of the section.

\begin{figure}
    \centering
    \includegraphics[width=0.48\textwidth]{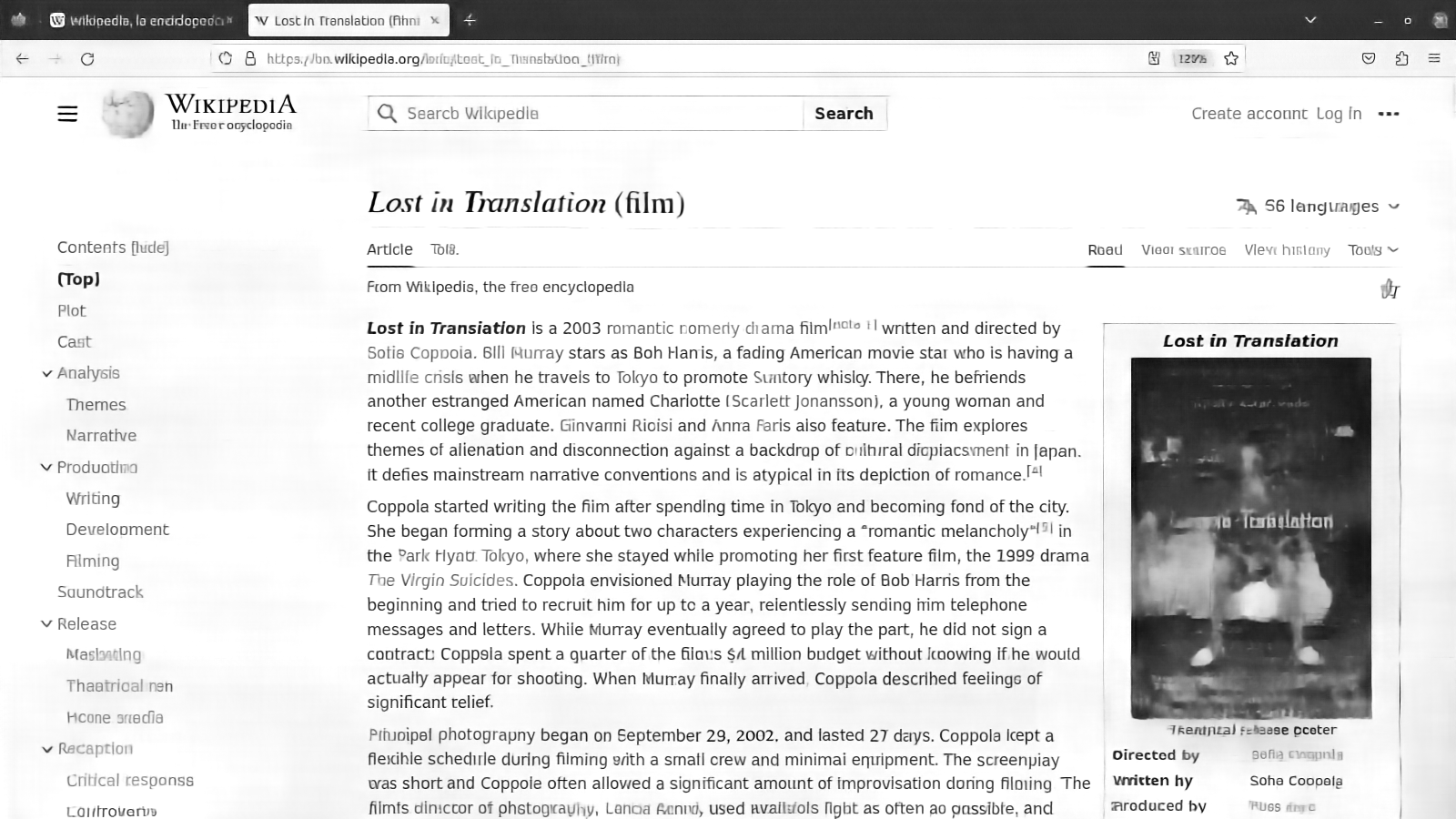}
    \caption{Example of a complete inference using the Pure Model in a real-life sample. }
    \label{fig:lost_in_translation}
\end{figure}

\begin{figure}
    \centering
    \includegraphics[width=0.4125\linewidth]{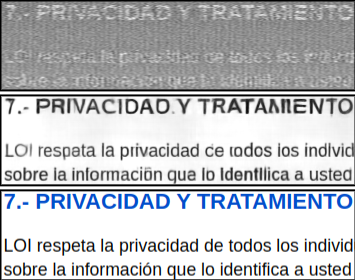}
    \includegraphics[width=0.57\linewidth]{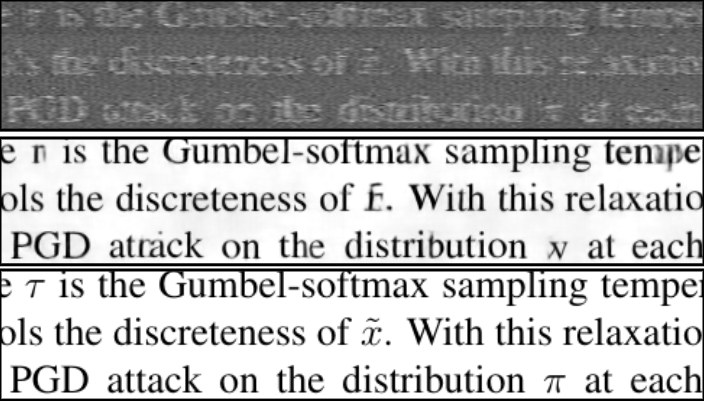}
    \caption{Zoomed-in examples obtained by vanilla \texttt{gr-tempest} (top), Pure Model (middle), and the original image (bottom). }
    
    \label{fig:resultados_end-to-end}
\end{figure}

The next step is, naturally, to re-train the model by using only real-life data. We will refer to the resulting system as the \emph{Pure Model}. Evaluation of its inferred images results in a CER of about 35\%, very similar to those obtained by the Base Models when evaluated on synthetic data. These are excellent results, which mean that only about one-third of the characters are incorrectly detected by Tesseract on the inferred image. Redundancy enables a human operator to recover most (if not all) of the rest of the text present in the image. A representative inference example is shown in Fig.~\ref{fig:lost_in_translation}.
Further zoomed-in results are shown in Fig.~\ref{fig:resultados_end-to-end}, including the results of vanilla \texttt{gr-tempest}. 
Note how, in the example on the left, the text is restored with higher quality when the font size is larger, even if the original text color is blue. Furthermore, the one on the right shows great text restoration performance except for some characters (such as ``$\tau$'', ``$\pi$'' and ``$\tilde x$'' symbols), which are less common and therefore under-represented in the training set.

\noindent \textbf{Denoising the grayscale images.} A pertinent question is how much information would have actually been lost had we not re-cast the TEMPEST problem as an inverse one. That is to say, what would the performance be had we proceeded as in~\cite{lemarchand2020electro,galvis2021denoising,long2024eye} and applied a denoiser to the grayscale image as produced by \texttt{TempestSDR} or \texttt{gr-tempest}. We have thus trained a model with only real-life signals as before, but taking the magnitude of the complex samples. This results in a significant increase in the CER, reaching almost 44\%. 
This shows that using the complex samples as an input to the network is a better choice, as the system can leverage information from both magnitude and phase.

\begin{table}[]
\small
\centering
\begin{tabular}{|cccc|}
\hline
\multicolumn{1}{|c|}{Model}        & \multicolumn{1}{c|}{PSNR (dB)} & \multicolumn{1}{c|}{SSIM} & CER (\%)\\ \hline \hline 

\multicolumn{4}{|c|}{Synthetic Data}\\ \hline
\multicolumn{1}{|c|}{Base (ideal pulse)} & \multicolumn{1}{c|}{\textbf{21.3}}     & \multicolumn{1}{c|}{\textbf{0.913}}  &   \textbf{29.5}  \\ \hline
\multicolumn{1}{|c|}{Base (real pulse)} & \multicolumn{1}{c|}{20.2}     & \multicolumn{1}{c|}{0.908}  &   32.8  \\ \hline \hline
\multicolumn{4}{|c|}{Real-life Data}\\ \hline
\multicolumn{1}{|c|}{Base (ideal pulse)} & \multicolumn{1}{c|}{10.0}     & \multicolumn{1}{c|}{0.610}  &   49.4  \\ \hline
\multicolumn{1}{|c|}{Base (real pulse)} & \multicolumn{1}{c|}{10.0}     & \multicolumn{1}{c|}{0.601}  &   55.2  \\ \hline
\multicolumn{1}{|c|}{\makecell{Raw image magnitude \\ (\texttt{gr-tempest})}} & \multicolumn{1}{c|}{8.57}     & \multicolumn{1}{c|}{0.345}  &   92.2  \\ \hline
\multicolumn{1}{|c|}{Pure (w/ complex values)}   & \multicolumn{1}{c|}{\textbf{15.2}}     & \multicolumn{1}{c|}{\textbf{0.787}} & \textbf{35.3} \\ \hline
\multicolumn{1}{|c|}{Pure (w/ magnitude only)}   & \multicolumn{1}{c|}{14.2}     & \multicolumn{1}{c|}{0.754} & 43.6 \\ \hline

\end{tabular}

\caption{Performance of all trained models, evaluated on test sets of both synthetic and real captures. The best performance for each dataset and metric is indicated in bold text.}
\label{table:evaluacion_test_metodos}
\end{table}

\noindent \textbf{On the utility of synthetic data.} As we discuss in the next section, robustness of the spying system requires signals that span several monitor configurations (i.e.\ resolutions) as well as SDR's parameters (i.e.\ harmonic and sampling rate). This means that the attacker has to build a training set including several thousands of real-life samples, which acquisition constitutes then a significant bottleneck in developing a robust spying system. It is crucial, then, to study how to reduce the number of real-life signals required and if it is possible to do so without affecting the resulting performance. 


The first idea is simply to build a smaller training set. 
For instance, if we use a third of the training set on the Pure Model, the CER would increase roughly by three percentage points, more precisely resulting in a CER of 38.3\%. 
Instead of training the Pure Model from scratch, a very interesting and useful alternative is to use the Base Model as a starting point, whose training samples are virtually free to produce. The idea is to expose the Base Model to real-life samples so that it can leverage what it has learned from the simulations to better infer images from real-life signals. 
More in particular, we start from the weights of the Ideal Base Model and further train it for another 100 epochs using only a subset of real-life samples. The results obtained with this methodology, a so-called Model Fine-Tuning (which may be interpreted as Few-Shot Learning in this case), is shown in Table \ref{table:fine_tuning}. Note how simulated data may be leveraged to obtain the same performance as the Pure Model but using only 10\% of the real-life samples. Quite interestingly, this fine-tuning produces the best results from all of the evaluated models. 

\begin{table}[]
\small
\centering
\begin{tabular}{|cccc|}
\hline
\multicolumn{1}{|c|}{Fraction}        & \multicolumn{1}{c|}{PSNR (dB)} & \multicolumn{1}{c|}{SSIM} & CER (\%)\\ \hline \hline 

\multicolumn{1}{|c|}{5\%} & \multicolumn{1}{c|}{14.6}     & \multicolumn{1}{c|}{0.766}  &   39.0  \\ \hline
\multicolumn{1}{|c|}{10\%} & \multicolumn{1}{c|}{15.2}     & \multicolumn{1}{c|}{0.791}  &   35.0  \\ \hline
\multicolumn{1}{|c|}{20\%} & \multicolumn{1}{c|}{15.4}     & \multicolumn{1}{c|}{0.797}  &   33.3  \\ \hline
\multicolumn{1}{|c|}{50\%} & \multicolumn{1}{c|}{15.6}     & \multicolumn{1}{c|}{0.803}  &   31.4  \\ \hline
\multicolumn{1}{|c|}{100\%} & \multicolumn{1}{c|}{15.7}     & \multicolumn{1}{c|}{0.806}  &   29.8  \\ \hline

\end{tabular}

\caption{Performance of the fine-tuned Base Model as we vary the number of real-life samples (as a fraction of the complete dataset) used in training. }
\label{table:fine_tuning}
\end{table}


\section{Robustness and Countermeasures}\label{sec:countermeasures_robustness}

\subsection{Robustness}\label{subsec:robustenss}

This section evaluates our system's performance when modifications are introduced in both the acquisition phase and the reference images. For instance, the training set was generated with a fixed sampling rate, tuning frequency, and monitor resolution configuration for both actual signals and simulations. It is essential to assess which changes in these parameters require complete retraining.

\begin{figure}
    \centering
    \subfloat[1600$\times$900 resolution image spied at 4th pixel frequency rate (CER = 26.6\%).\label{subfig:1600x900_harm4}]{\includegraphics[width=0.49\linewidth]{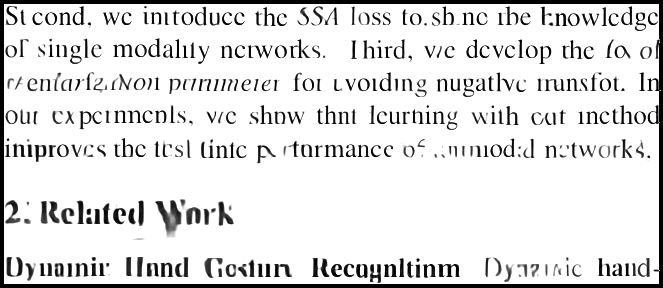}}
    \,
    \subfloat[1280x720 resolution image spied at 4th pixel frequency rate (CER = 50\%).\label{subfig:1280x720_harm4}]{\includegraphics[width=0.49\linewidth]{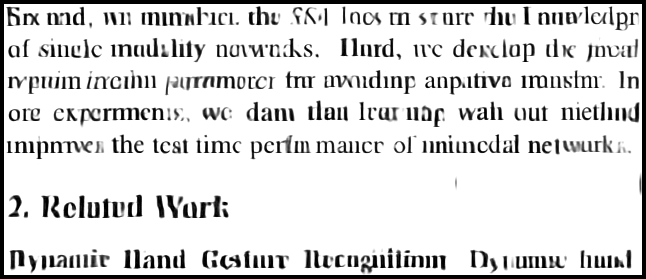}}
    
    \caption{Model inferences over non-trained setup spied images. The inference of \ref{subfig:1280x720_harm4} shows the model does not assure a good performance at other spying setups.}

    \label{fig:inferencias_otras_configuraciones}
\end{figure}

\noindent \textbf{Robustness to the Signal Acquisition Process}
We start by exploring changes in SDR tuning frequency. Our choice of the third-pixel harmonic was based on the absence of other significant sources of radio-frequency interference, but this is not always the case, and the operator may need to tune to, for instance, the fourth one. Note that in this case, the most important difference between the samples in the training set and the observed signal lies in the form of $g(t)$ (cf.\ Eq.~\eqref{eq:senial_capturada}), which will now correspond to another $f_c$. However, as illustrated in Fig.~\ref{fig:ejemplo_G}, the difference in the corresponding pulses is not significant, and the learning system should obtain reasonable results. This is confirmed in Fig.~\ref{subfig:1600x900_harm4}, which shows an inference example using a real signal tuned at $f_c=4/T_p$. The resulting CER in this example was 26\%, demonstrating the robustness to changes in the tuning frequency. 

As a second step, let us additionally modify the monitor's resolution (thus resulting in a different pixel rate $1/T_p$) and choose again $f_c=4/T_p$. We interpolated the captured complex image resolution to 1600$\times$900 before computing the inference to feed the learning module with the same array size that it was trained on, thus avoiding any disadvantage compared to the previous configuration. An example inference (using $1280\times 720@60$fps) is shown in Fig. \ref{subfig:1280x720_harm4}. In this case, the performance was clearly degraded, resulting in a CER of 50\%. Differently from the previous case, differences in the resulting shaping pulses are enough to produce samples where the learning system's performance degrades significantly. 

In any case, expecting the system to perform well under all possible resolutions and harmonics would not be reasonable. However, since the number of possible configurations is limited, we may envisage a set of different parameters for the DRUNet, each trained on signals acquired when a specific resolution was used in the monitor and a certain configuration was used on the SDR. 
As discussed in the previous section, we may fine-tune the model trained on simulations, so the acquisition process should not be time-consuming.


\noindent \textbf{Robustness to the Images' Content}
Text fonts not used for training are another point to consider for testing the model's robustness, appearing in the examples we showed previously, especially that of Fig.~\ref{fig:resultados_end-to-end}. Given that several of the images we included in our dataset come from PDF documents obtained from a conference (and thus with the same font), it is interesting to evaluate whether the system presents certain overfitting to these kinds of images. 
To measure the performance of the model for unseen fonts, we created a new dataset consisting of 800 new simulated samples. Each of these images consists of random text, where each line alternates between 147 different font types (those included in the default Ubuntu installation and that contain the Latin script). The simulation uses the same image resolution, pixel harmonic frequency, and sampling rate as in the previous section. 



Using a subset of 300 of these images to evaluate the Base Model with the ideal pulse results in an increase of the average CER, that moves from about the 30\% that we obtained before (cf.\ Table \ref{table:evaluacion_test_metodos}) to 48.7 \%. However, simply by further training the model for another 10 epochs, where the remaining 500 samples were added to the training set, the resulting CER drops again to 29.8 \%. This experiment shows that the architecture has the potential to learn new text font types with a few training epochs and provides further evidence of its expressiveness.

\subsection{Countermeasures}\label{subsec:countermeasures}

It is essential to expose the spying system flaws so the counterpart (e.g., the computer user) can exploit them and ensure the protection of personal or classified information. To this end, we mention two countermeasures that, by modifying the displayed image (in a primarily eye-imperceptible manner to the computer user), inference based on the resulting emanations fails. These defects stem from the analysis discussed in Sec.~\ref{sec:hdmi} and leverage the non-linearity of the TMDS encoding.

\begin{figure}
\centering


\includegraphics[width=0.49\linewidth]{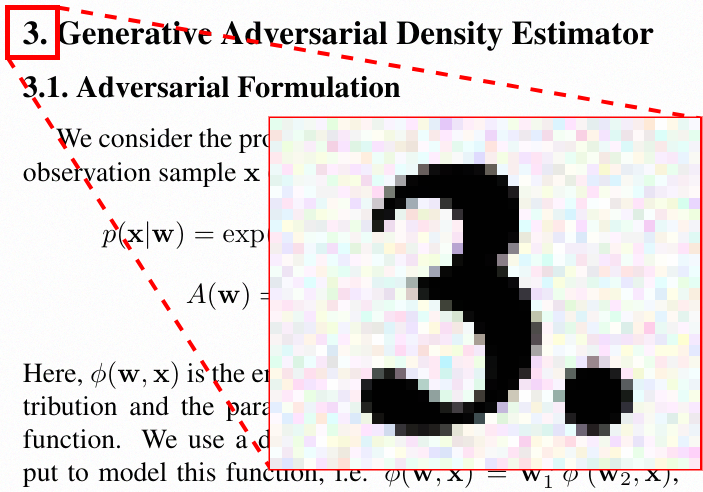}
\includegraphics[width=0.49\linewidth]{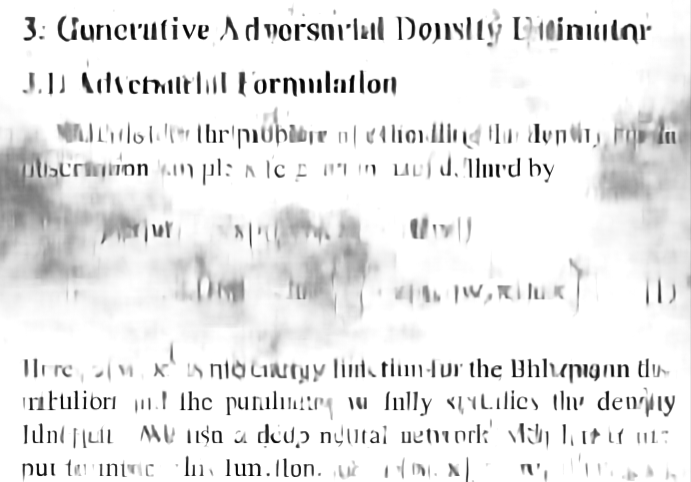}
\caption{Image inferences when synthetic low-level noise is added to the original image. Inference performance is significantly degraded, even with an imperceptible noise level.}
\label{fig:protección_con_ruido_3std}
\end{figure}

One way to accomplish this is by adding low-level noise to the image displayed on the monitor, creating an adversarial attack on the neural network. This noise may be, for instance, an additive Gaussian noise with a constant variance. The example in Fig.~\ref{fig:protección_con_ruido_3std} illustrates this possibility by artificially adding a very small noise to the original image ($\sigma=3$). Note how most of the text in the inference becomes illegible.

\begin{figure}
    
    \centering
    \subfloat[Image with horizontal gradient.\label{subfig:gradient_image}]{\includegraphics[trim={65px 180px 1020px 540px}, clip, width=0.4925\linewidth]{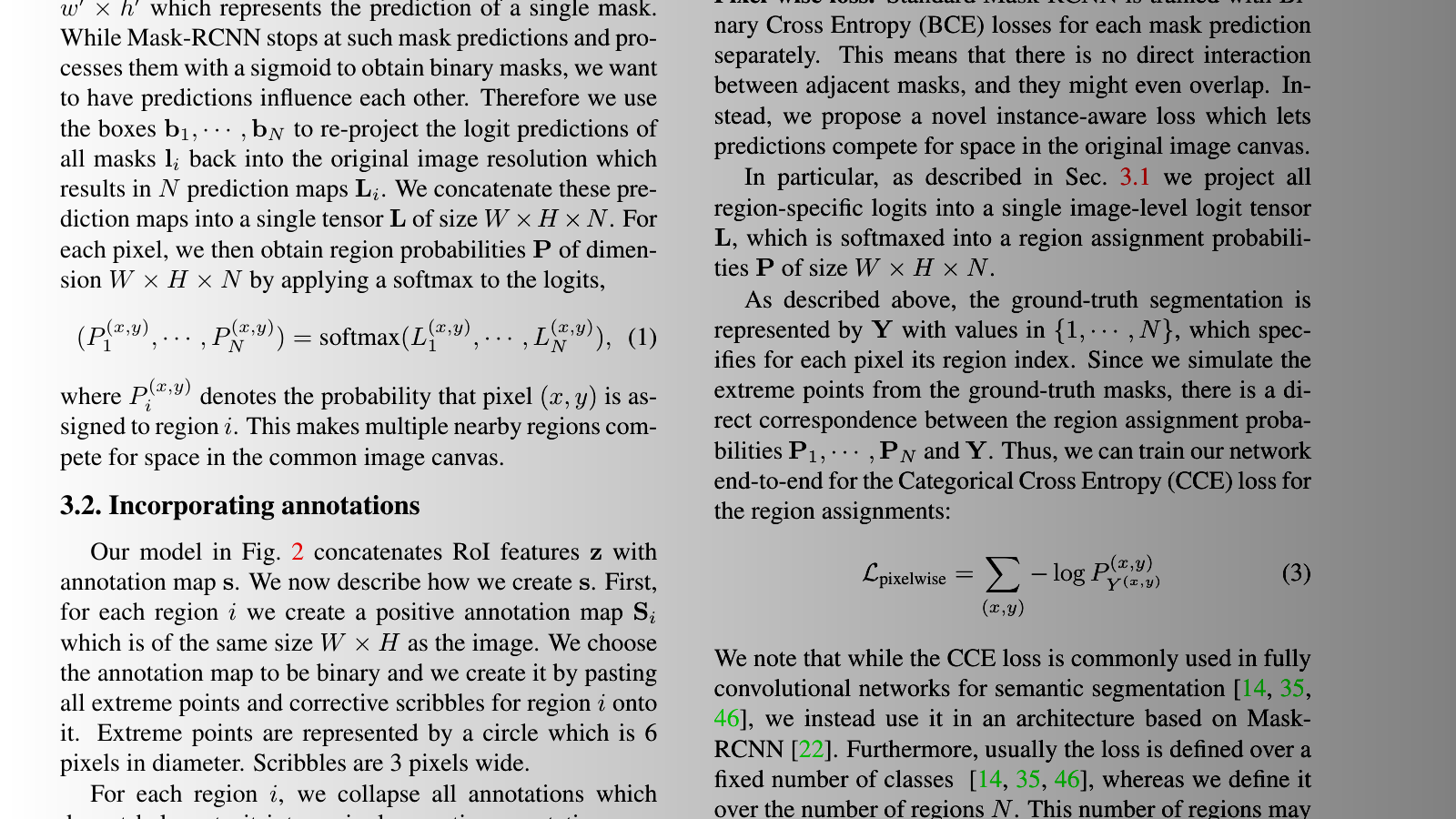}}
    \,    
    \subfloat[Eavesdropped image.\label{subfig:gradient_image_capture_inference}]{\includegraphics[trim={0px 0px 160px 0px}, clip, width=0.4925\linewidth]{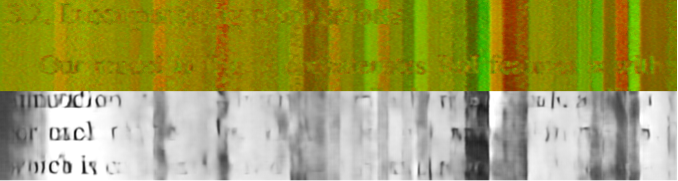}}
    \caption{Gradient background experiment scenario. A horizontal 0-127 grayscale ramp is subtracted from the original image (a), resulting in an observed complex image (upper b) with several vertical bands. Inference (lower b) thus fails to restore the text.
    }

    \label{fig:gradient_effect}
\end{figure}

A more perceptible but definitive solution is to use a color gradient on the images' background, as illustrated in Fig.~\ref{subfig:gradient_image}. When using a horizontal gradient (a white-to-black ramp, for example), we are changing the grayscale linearly over the image, but the TMDS encoding will produce significant changes on the eavesdropped signal (see Fig.~\ref{subfig:gradient_image_capture_inference}). In this case, also shown in Fig.~\ref{subfig:gradient_image_capture_inference}, the inference fails completely.

\section{Conclusion}\label{sec:conclusions}
In this work, we have presented an open-source implementation of a deep learning architecture trained to map from the electromagnetic signal emanating from an HDMI cable to the displayed image. The complete dataset, including simulations based on the analytical expressions we derived (as well as scripts to generate them), is also made available. Notably, the system obtains much better results than previous implementations, significantly improving the Character Error Rate when eavesdropping text. 

This work paves the way for several interesting and challenging research avenues. As we discussed in Sec.~\ref{sec:countermeasures_robustness}, the trained architecture's performance degrades as we modify the spied system's parameters (e.g.,~the resolution or the tuned frequency). A possible solution is to train several architectures, one for each foreseeable set of parameters. Simulations will naturally come in handy in this otherwise extremely time-consuming process. An alternative is to leverage the fact that we have an explicit expression for the degradation operator and strive at solving~\eqref{map_equation} directly. Deep learning has also been successfully applied to these so-called plug\&play methods, in particular, to apply the prior distribution or regularization term, which takes the form of a denoiser (see~\cite{zhang2017learning} for example). 
The main challenge in the case of TEMPEST is how to efficiently find the optimum to the data term since the degradation operator is highly non-linear.

We may also enrich the signal we are using for inference. As we discussed before, the eavesdropped samples present significant redundancy, which we implicitly used through \texttt{gr-tempest} to align $\bm{Y}$ and $\bm{X}$. However, this redundancy may also be used to produce even better results. We may, for instance, use several consecutive complex arrays of samples to construct a complex tensor, which may then be fed to a network that infers the original image. 

Finally, it is important to highlight that the architecture we used takes some seconds to produce each inference. This is hardly real-time, and it would be interesting to undertake a faster implementation now that the method's feasibility has been verified. 


\bibliographystyle{ACM-Reference-Format}
\bibliography{biblio}

\end{document}